\documentclass[11pt,letterpaper]{article}

\usepackage[
  letterpaper,
  top=1.222in,
  right=1.417in,
  bottom=1.972in,
  left=1in,
  headheight=0pt,
  headsep=0pt,
  footskip=0.49in
]{geometry}

\usepackage[T1]{fontenc}
\usepackage[utf8]{inputenc}
\usepackage{libertine}
\usepackage{amsmath,amssymb,amsthm}
\usepackage[libertine]{newtxmath}
\usepackage[scaled=0.92]{inconsolata}
\usepackage{microtype}
\usepackage{graphicx}
\usepackage{booktabs,tabularx,array}
\usepackage{float}
\usepackage{enumitem}
\usepackage{caption}
\usepackage{titlesec}
\usepackage{listings}
\usepackage{xcolor}
\usepackage{hyperref}
\hypersetup{hidelinks}
\graphicspath{{figures/}}

\newif\ifanonymousreview
\anonymousreviewfalse

\AtBeginDocument{\fontsize{9}{13.5}\selectfont}
\titleformat{\section}
  {\normalfont\biolinum\bfseries\fontsize{9}{11.25}\selectfont}
  {\thesection}{0.55em}{\MakeUppercase}
\titlespacing*{\section}{0pt}{10pt}{2pt}

\titleformat{\subsection}
  {\normalfont\biolinum\bfseries\fontsize{9}{11.25}\selectfont}
  {\thesubsection}{0.55em}{}
\titlespacing*{\subsection}{0pt}{7pt}{2pt}

\titleformat{\subsubsection}
  {\normalfont\itshape\fontsize{9}{11.25}\selectfont}
  {\thesubsubsection}{0.55em}{}
\titlespacing*{\subsubsection}{0pt}{6pt}{1.5pt}

\DeclareCaptionFont{icvrtcaption}{\biolinum\fontsize{8}{10}\selectfont}
\setlist[itemize]{leftmargin=18pt,itemsep=0pt,topsep=3pt,parsep=0pt}
\setlist[enumerate]{leftmargin=20pt,itemsep=0pt,topsep=3pt,parsep=0pt}

\makeatletter
\newcommand{\@papertitle}{Paper Title}
\newcommand{\@shorttitle}{}
\newcommand{\@papersubtitle}{}
\newcommand{\@paperauthors}{}

\newcommand{\papertitle}[1]{\gdef\@papertitle{#1}}
\newcommand{\shorttitle}[1]{\gdef\@shorttitle{#1}}
\newcommand{\papersubtitle}[1]{\gdef\@papersubtitle{#1}}
\newcommand{\paperauthors}[1]{\gdef\@paperauthors{#1}}

\newcommand{\authorentry}[4]{
  {\biolinum\fontsize{11}{16}\selectfont\MakeUppercase{#1}\par}
  {\fontsize{8}{10}\selectfont
    \href{https://orcid.org/#2}{ORCID: #2}\par}
  \vspace{2pt}
  {\fontsize{9}{14.85}\selectfont #3, \href{mailto:#4}{#4}\par}
}
\newcommand{\authorentrynoorcid}[3]{
  {\biolinum\fontsize{11}{16}\selectfont\MakeUppercase{#1}\par}
  \vspace{2pt}
  {\fontsize{9}{14.85}\selectfont #2, \href{mailto:#3}{#3}\par}
}
\newcommand{\authornote}[1]{
  \vspace{5pt}
  {\fontsize{8}{11}\selectfont #1\par}
}

\newcommand{\makepapertitle}{
  \begingroup
  \raggedright
  {\biolinum\bfseries\fontsize{12}{18}\selectfont\@papertitle\par}
  \ifx\@shorttitle\@empty\else
    {\fontsize{10}{15}\selectfont\@shorttitle\par}
  \fi
  \ifx\@papersubtitle\@empty\else
    {\biolinum\fontsize{9}{16.65}\selectfont\@papersubtitle\par}
    \vspace{18pt}
  \fi
  \ifanonymousreview\else
    \@paperauthors
  \fi
  \endgroup
  \vspace{5pt}
}
\makeatother

\renewenvironment{abstract}
  {\par\vspace{6pt}\begingroup\fontsize{8}{12}\selectfont\noindent\ignorespaces}
  {\par\endgroup}

\newcommand{\ccsconcepts}[1]{
  \par\vspace{4pt}\noindent\textbf{CCS CONCEPTS}\enspace #1\par
}
\newcommand{\keywords}[1]{
  \par\vspace{4pt}\noindent\textbf{Additional Keywords and Phrases:} #1\par
}

\theoremstyle{plain}

\let\oldthebibliography\thebibliography
\let\endoldthebibliography\endthebibliography
\renewenvironment{thebibliography}[1]
  {\begingroup\fontsize{7}{8}\selectfont
   \oldthebibliography{#1}
   \setlength{\itemsep}{0pt plus 0.3pt}\setlength{\parskip}{0pt}}
  {\endoldthebibliography\endgroup}

\papertitle{Big World, Little Me: Comparing a First-Person VR Walkthrough Video
  and Static Slides for Communicating Height-Related Accessibility Barriers}

\papersubtitle{}
\paperauthors{
  \authorentry{Chan In Sio}{0009-0002-2701-5424}
    {The Hong Kong Polytechnic University}{devin.sio@connect.polyu.hk}
  \authorentry{Alex Mann}{0009-0002-4248-3779}
    {The Hong Kong Polytechnic University}{alex.mann@connect.polyu.hk}
  \authorentry{Qian Ming Jia}{0009-0000-9810-6919}
    {The Hong Kong Polytechnic University}{kitty.jia@connect.polyu.hk}
  \authorentrynoorcid{Xuecheng Guo}
    {The Hong Kong Polytechnic University}{24073862g@connect.polyu.hk}
  \authorentrynoorcid{Yan Zhuo}
    {The Hong Kong Polytechnic University}{24121182g@connect.polyu.hk}
  \authorentrynoorcid{Xinyu Ning}
    {The Hong Kong Polytechnic University}{24118592g@connect.polyu.hk}
  \authorentrynoorcid{Kaikuo Liao}
    {The Hong Kong Polytechnic University}{24092803g@connect.polyu.hk}
  \authorentry{Lik-Hang Lee\textsuperscript{*}}{0000-0003-1361-1612}
    {The Hong Kong Polytechnic University}{lik-hang.lee@polyu.edu.hk}
  \authornote{\textsuperscript{*}Corresponding author: Lik-Hang Lee.}
}

\begin{document}

\makepapertitle

\hypersetup{
  pdftitle={Big World, Little Me: Comparing a First-Person VR Walkthrough Video
    and Static Slides for Communicating Height-Related Accessibility Barriers},
  pdfauthor={Chan In Sio; Alex Mann; Qian Ming Jia; Xuecheng Guo; Yan Zhuo;
    Xinyu Ning; Kaikuo Liao; Lik-Hang Lee},
  pdfsubject={VR-authored accessibility narratives},
  pdfkeywords={virtual reality, accessibility, dwarfism, perspective-taking,
    narrative engagement, state empathy}
}

\begin{abstract}
People with dwarfism and other forms of short stature may encounter barriers
created by fixed-height environments and discriminatory social assumptions.
This paper evaluates \emph{Big World, Little Me}, a 10~min 52~s first-person
walkthrough recorded from a Unity-based VR environment. The narrative depicts
domestic reachability, public-transport, and employment barriers. In a
non-randomised between-subjects study, 100 participants aged 15--48 years viewed
either the fixed-duration audiovisual walkthrough or a silent, self-paced
27-slide
presentation representing the same principal events.
Narrative engagement was higher for the walkthrough ($M=5.54$, $SD=0.51$)
than for the slides ($M=4.24$, $SD=0.73$), mean difference $=1.30$, 95\% CI
[1.05, 1.55], Hedges' $g=2.05$, $p<.001$. State empathy was also higher for
the walkthrough ($M=3.15$, $SD=0.33$) than for the slides ($M=2.21$,
$SD=0.60$), mean difference $=0.94$, 95\% CI [0.74, 1.13], Hedges'
$g=1.91$, $p<.001$. Exploratory supportive-intention ratings increased more
in the walkthrough condition than in the slide condition. Participants also
rated the walkthrough as more dynamic, continuous, spatially clear, and
information-rich. The findings indicate stronger immediate responses to the
complete walkthrough format, but do not isolate the effects of motion,
continuity, pacing, information quantity, or participant control. The
walkthrough should be understood as screen-based VR-authored media rather than
immersive VR or a reproduction of lived experience.
\end{abstract}

\ccsconcepts{\textbullet\ Human-centered computing $\rightarrow$ Virtual reality
  \textbullet\ Human-centered computing $\rightarrow$ Empirical studies in HCI
  \textbullet\ Human-centered computing $\rightarrow$ Accessibility technologies}

\keywords{virtual reality, accessibility, dwarfism, perspective-taking,
  narrative engagement, disability representation, user study}

\section{Introduction}

People with dwarfism and other forms of short stature may face barriers created
by environments and social practices designed around average-height users.
Fixed-height facilities can restrict everyday activities, while infantilising
assumptions and appearance-based judgments can affect social participation and
employment. This reflects an interactional view of disability, in which
participation is shaped by environmental and attitudinal conditions rather than
impairment alone~\cite{who2023disability}. Research on achondroplasia also
reports limitations in physical functioning and psychosocial quality of life
across the lifespan~\cite{constantinides2022achondroplasia}. Public communication
should therefore highlight structural and social barriers, not only medical
characteristics.

Virtual reality (VR) may make such barriers more concrete by presenting events
from a first-person viewpoint and preserving spatial relationships between the
viewer, surrounding objects, and other people. Prior perspective-taking studies
have reported some positive attitudinal, prosocial, and cognitive-empathy
effects, although findings were not consistently stronger across all
outcomes~\cite{herrera2018empathy,vanloon2018empathy}. VR should therefore not
be treated as an automatic ``empathy machine''~\cite{soradomenjo2022empathy}.

This caution is particularly important in disability-related work. Brief
simulations may imply that temporary inconvenience reproduces lived experience,
while foregrounding helplessness and overlooking adaptation, agency, community,
and discrimination. Disability simulations can increase distress and perceived
vulnerability without reliably improving attitudes or willingness to
interact~\cite{narioredmond2017crip}. Accordingly, this study presents one
fictionalised accessibility narrative intended to direct attention toward
environmental and social barriers, not to reproduce what it is like to live
with dwarfism.

The study builds on \emph{Big World, Little Me}, a Unity-based VR environment
developed for Meta Quest 3. It allows users to follow selected everyday
situations from the first-person perspective of an adult protagonist with
dwarfism. The participant-facing narrative comprised three connected settings:
domestic activities involving a fixed-height sink and refrigerator; public
transport involving inaccessible controls and being mistaken for a child; and
job interviews involving height-related assumptions and rejection.

Because HMD-based VR requires specialised hardware and setup, we evaluated a
recorded first-person walkthrough as a more accessible presentation format.
Although video removes interaction, head tracking, stereoscopic depth, and
embodiment, it preserves the first-person viewpoint, narrative sequence, and
dynamic spatial information of the VR environment.

We compared a 10~min 52~s walkthrough of \emph{Big World, Little Me} with a
corresponding self-paced static-slide presentation in a non-randomised between-subjects
study with 100 participants. We examined whether presentation format was
associated with narrative engagement and state empathy, while treating
supportive intentions as exploratory. The study compared the two complete
presentation formats and was not designed to establish a causal mediation
pathway between engagement, empathy, and intentions.

\begin{center}
  \setlength{\fboxsep}{6pt}
  \setlength{\fboxrule}{0.5pt}
  \fcolorbox{black!55}{black!4}{
    \begin{minipage}{0.94\linewidth}
      {\biolinum\bfseries Research Questions\par}
      \vspace{3pt}

      \textbf{RQ1:} Do participants in the walkthrough-video condition report
      greater narrative engagement than participants in the static-slide
      condition?

      \smallskip
      \textbf{RQ2:} Do participants in the walkthrough-video condition report
      greater state empathy than participants in the static-slide condition?

      \smallskip
      \textbf{RQ3:} How do supportive intentions differ between the two
      presentation conditions?
    \end{minipage}
  }
\end{center}

This paper makes three contributions. First, it describes how a Unity-based
first-person accessibility experience was adapted into a recorded walkthrough
and a corresponding self-paced static-slide presentation. Second, it presents a non-randomised
between-subjects comparison of the two formats using measures of narrative
engagement and state empathy. Third, it offers preliminary practical insights
for interpreting screen-based VR-authored narratives, particularly where the
walkthrough differs from static slides in both dynamic presentation and
perceived information.

\section{Related Work}

\subsection{VR Perspective-Taking and Narrative Engagement}

VR perspective-taking has been used to make unfamiliar social situations more
concrete through situated, first-person representation. In studies of
homelessness and social bias, immersive VR produced some positive effects on
attitudes, supportive behaviour, cognitive empathy, and hiring decisions, but
these effects were not consistent across all outcomes
\cite{herrera2018empathy,vanloon2018empathy,crone2022avatar}. These mixed
findings caution against treating immersion as an automatic route to empathy.

One possible explanation lies in narrative engagement rather than immersion
alone. Busselle and Bilandzic describe narrative engagement through narrative
understanding, attentional focus, emotional engagement, and narrative presence
\cite{busselle2009engagement}. Green and Brock similarly link narrative
transportation with attention, imagery, affect, and story-consistent judgments
\cite{green2000transportation}. Dynamic presentation may support these processes
when movement and temporal order are important: instructional animations often
outperform static pictures under such conditions
\cite{hoeffler2007animation}. However, richer media can also increase presence
and affect without improving learning, and may introduce greater cognitive load
\cite{parong2018learning,makransky2019presence}. These findings motivate direct
measurement of engagement and immediate empathy rather than assuming that a
more dynamic format is necessarily more effective.

\subsection{Disability Representation and Height-Related Barriers}

Perspective-taking involving disability requires additional ethical caution. A
brief simulation cannot reproduce a lifelong and socially situated experience,
and may foreground helplessness or vulnerability while overlooking adaptation,
agency, community, and structural discrimination. Temporary disability
simulations have been shown to increase distress and perceived vulnerability
without reliably improving attitudes or willingness for future interaction
\cite{narioredmond2017crip}. Disability-studies-informed HCI therefore cautions
against treating disability solely as individual deficit and stresses the value
of disabled people's knowledge in design~\cite{mankoff2010disability}.
Disability-inclusive co-design has been proposed as an important approach for
incorporating disabled people's perspectives into technology research and
design~\cite{bandukda2022codesign}.

This concern is directly relevant to people with dwarfism and other forms of
short stature. Research on achondroplasia and skeletal dysplasias identifies
barriers across physical functioning, mobility, self-care, household tasks,
public transport, employment, access to services, and social participation
\cite{gollust2003achondroplasia,sommer2017aples,hyvonen2020functioning}.
Discrimination and underestimation by others further show that these
difficulties arise from both environmental design and social attitudes.
Accordingly, the present work frames \emph{Big World, Little Me} as one
fictionalised narrative about modifiable barriers, not as an opportunity for
viewers to temporarily ``become'' a person with dwarfism.

\subsection{From Immersive VR to Screen-Based Walkthroughs}

HMD-based VR can support embodiment and sensorimotor interaction, and altered
virtual bodies have influenced social judgments in some studies
\cite{yee2006digital,peck2013avatar,banakou2016embodiment}. These effects depend
on features such as tracked viewpoint, body ownership, agency, and
sensorimotor contingencies that are absent from conventional video. HMD
delivery also requires dedicated hardware and setup and may introduce
locomotion discomfort, for which field-of-view restriction is one established
mitigation approach \cite{fernandes2016sickness,alzayer2019fov}.

Recorded walkthroughs offer a more scalable way to distribute VR-authored
content through classrooms, online platforms, and public communication. They
retain first-person viewpoint, temporal sequence, dynamic movement, and spatial
relationships, while removing stereoscopic depth, head tracking, embodiment,
and interaction. The present study therefore evaluates screen-based
VR-authored media rather than immersive VR. It compares a recorded walkthrough
with corresponding static slides to examine whether participants in the two
conditions report different levels of narrative engagement and state empathy.

\section{Design of the VR-Authored Stimuli}

\subsection{VR Environment and Narrative Design}

\emph{Big World, Little Me} was developed in Unity as a first-person VR
environment for Meta Quest 3. The experience follows an adult protagonist with
dwarfism through selected everyday situations and focuses on environmental and
social barriers created by average-height assumptions. Rather than presenting
dwarfism primarily as a medical condition, the narrative highlights how
physical design, public interactions, and employment practices can restrict
participation.

The participant-facing narrative comprised three connected settings. In the
domestic setting, the protagonist moved a chair to reach a fixed-height sink and
attempted to retrieve food from a refrigerator. In the public-transport setting,
the protagonist encountered inaccessible controls and was mistaken for a child.
In the employment setting, the protagonist attended job interviews in which
recruiters made height-related assumptions about physical capacity, teamwork,
and employability.

\begin{figure}[!htbp]
  \centering
  \includegraphics[width=0.9\linewidth]{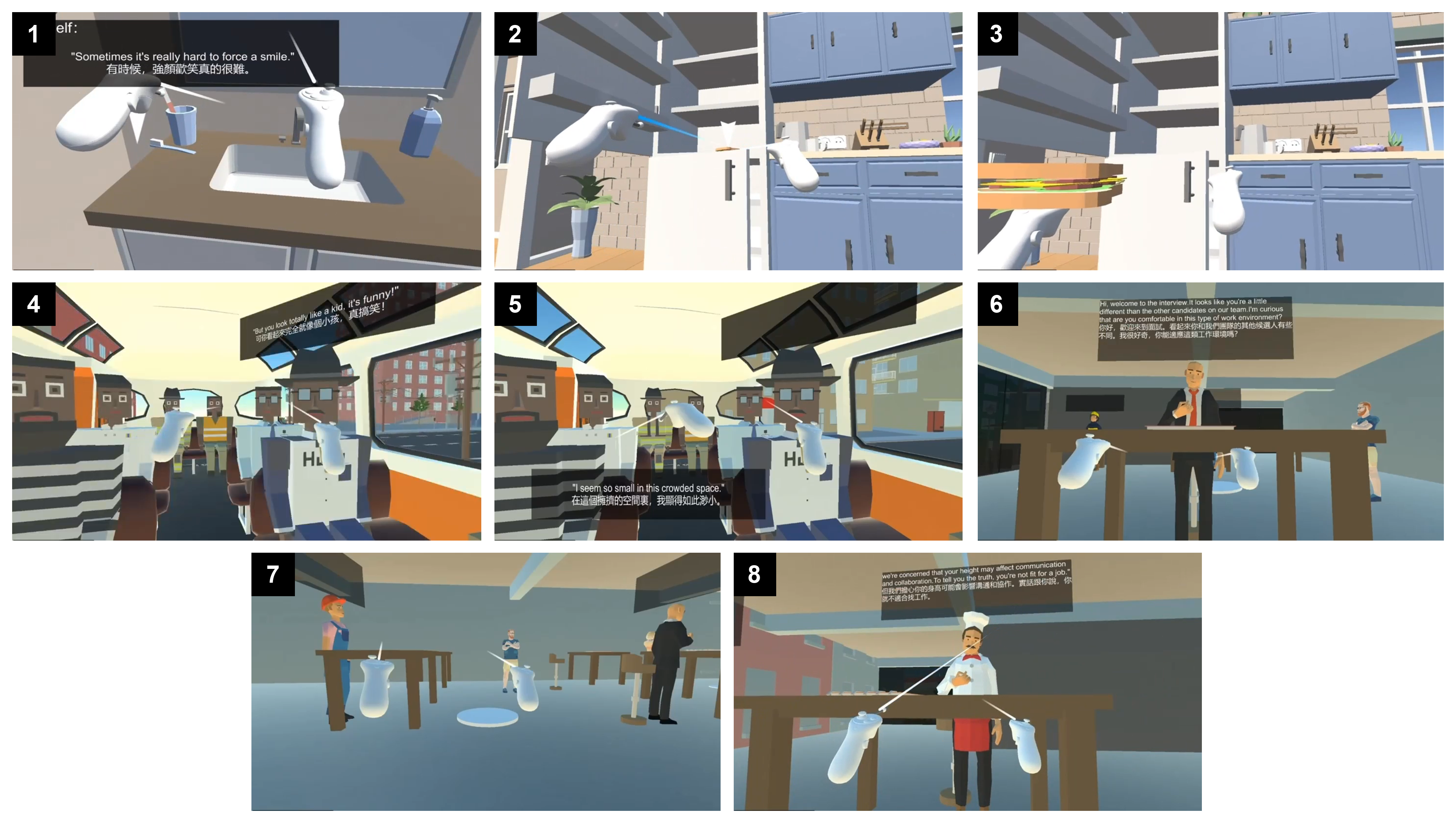}
  \caption{Representative frames from the first-person walkthrough video. The
  walkthrough shows (1--3) domestic reachability and object-access barriers,
  including use of a fixed-height bathroom sink and retrieval of food from a
  refrigerator; (4--5) public transport and an interaction in which the adult
  protagonist is mistaken for a child; and (6--8) job-market encounters
  involving repeated height-related assumptions about capacity and
  employability. Controller representations, interaction rays, and dialogue
  boxes are elements of the Unity environment.}
  \label{fig:vr-narrative}
\end{figure}

The VR environment included first-person movement, object manipulation,
controller-based interaction, trigger zones, character dialogue, subtitles, and
transitions between settings. Representative views of the three
participant-facing settings are shown in
Fig.~\ref{fig:vr-narrative}.

\subsection{First-Person Walkthrough Video}

We created a 10~min 52~s first-person walkthrough video as a scalable,
non-HMD presentation of the VR environment. The recording retained the
protagonist's first-person viewpoint, controller-based interactions, character
dialogue, subtitles, environmental audio, and transitions across the domestic,
public-transport, and employment settings. This viewpoint preserved the visible
spatial relationships between the protagonist and fixed-height objects,
transport fixtures, furniture, and other characters.

The walkthrough was captured from the display of a PC connected to the VR
system and exported at a resolution of \(960 \times 528\) pixels and a frame
rate of 20~fps. Editing was limited to removing loading screens, prolonged
pauses, and development artefacts. No new narrative events were introduced,
and the original event order was retained. Participants viewed the video on a
conventional two-dimensional display rather than through an HMD. The condition
is therefore described as a \emph{first-person VR walkthrough video}, not as
immersive or embodied VR. The presentation also stated that it represented one
fictionalised account of selected accessibility barriers rather than a
reproduction of lived experience.

\subsection{Static-Slide Presentation}

The comparison condition was a 27-slide presentation constructed from
screenshots of the same domestic, public-transport, and employment settings.
The slides followed the walkthrough's narrative order and represented the
principal events, including reaching the sink, retrieving food from the
refrigerator, travelling by bus, being mistaken for a child, and attending job
interviews.

The slide deck was designed to correspond to the walkthrough in event order,
written dialogue, and subtitle content. Multi-step actions were represented
through successive still frames, while narration, environmental audio,
continuous movement, object animation, character motion, and environmental
transitions were absent. Participants viewed the presentation in full-screen
mode and advanced through the slides at their own pace.

\begin{figure}[!htbp]
  \centering
  \includegraphics[width=0.9\linewidth]{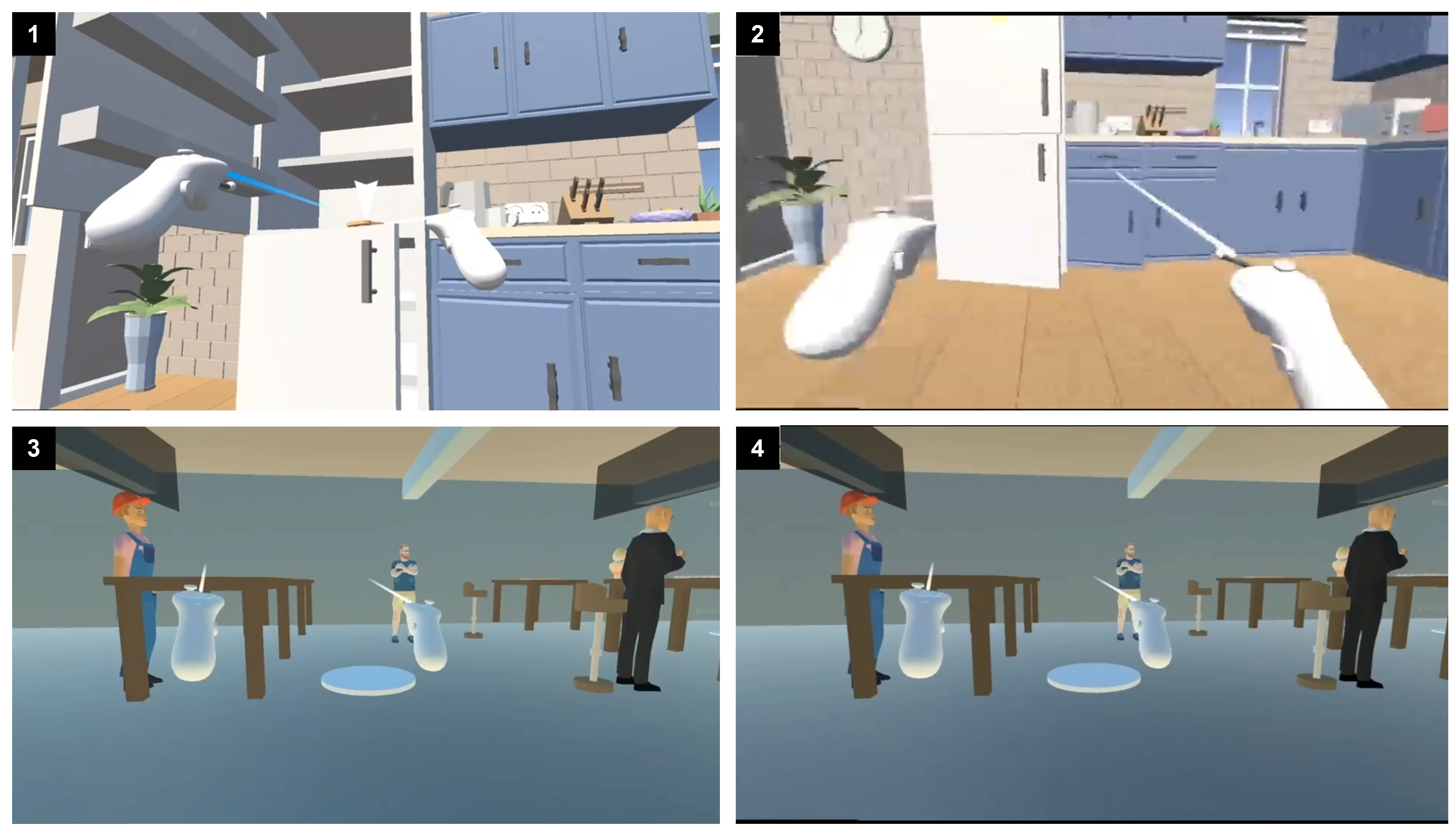}
  \caption{Walkthrough-video and static-slide presentation formats. Panels 1 and
  3 show frames from the continuous refrigerator and employment sequences.
  Panels 2 and 4 show the corresponding participant-facing slides. Written
  dialogue, subtitle content, and event order were designed to correspond across
  conditions, whereas narration, environmental audio, continuous movement, and
  environmental transitions were present only in the walkthrough-video
  condition.}
  \label{fig:format-comparison}
\end{figure}

Figure~\ref{fig:format-comparison} illustrates corresponding domestic and
employment events in the two conditions. Panels 2 and 4 show the corresponding
participant-facing slides.

\subsection{Stimulus Correspondence}

Both conditions presented the same domestic, public-transport, and employment
settings in the same narrative order, using corresponding first-person visual
material, dialogue, and subtitles. The walkthrough presented these events
through continuous movement, object and character animation, and environmental
transitions over a fixed duration of 10~min 52~s. The slide condition represented
the same principal events through 27 sequential still frames that participants
advanced at their own pace.

Both conditions were displayed in full-screen mode. However, they differed in
movement, temporal continuity, pacing, participant control, narration, and
environmental audio. The experiment therefore compared two complete
presentation formats rather than isolating the effect of motion or any other
single feature. Both presentations framed the material as one fictionalised
account of selected accessibility barriers rather than a universal
representation of living with dwarfism.

\section{User Study}

\subsection{Study Design and Hypotheses}

We conducted a two-group, non-randomised between-subjects study comparing a
first-person VR walkthrough video with a static-slide presentation. Participants
viewed only one format. The first 50 participants completed the walkthrough-video
condition, followed by 50 participants who completed the static-slide condition.
Condition allocation was therefore determined by recruitment phase rather than
random assignment.

Narrative engagement was the primary outcome, and state empathy was the
secondary outcome. Both were measured immediately after stimulus exposure using
validated scales. Supportive behavioural intentions were examined
exploratorily using researcher-developed items.

\begin{center}
  \setlength{\fboxsep}{6pt}
  \setlength{\fboxrule}{0.5pt}
  \fcolorbox{black!55}{black!4}{
    \begin{minipage}{0.94\linewidth}
      {\biolinum\bfseries Hypotheses\par}
      \vspace{3pt}

      \textbf{H1:} Participants in the walkthrough-video condition would report
      greater narrative engagement than participants in the static-slide
      condition.

      \smallskip
      \textbf{H2:} Participants in the walkthrough-video condition would report
      greater state empathy than participants in the static-slide condition.
    \end{minipage}
  }
\end{center}

The study compared two complete presentation formats. Because participants were
not randomly assigned, the findings indicate between-group associations and do
not establish that presentation format alone caused the observed differences.

\subsection{Participants}

Participants were recruited in Hong Kong through electronic advertisements and
participant referrals. Eligible participants were required to have sufficient
English proficiency to understand the materials and to be able to view the
assigned presentation on an iPad. Participants under 18 years
of age took part with parent or guardian consent and provided their own assent.
Individuals with uncorrected visual or hearing difficulties that prevented
access to the stimuli were not eligible.

The final sample comprised 100 participants, with 50 participants in each
condition. Participants were aged between 15 and 48 years ($M=23.51$,
$SD=8.71$). Mean age was similar between the walkthrough ($M=23.12$,
$SD=8.60$) and slide ($M=23.90$, $SD=8.89$) conditions. The respective
numbers of participants under 18 were 20 and 15. The sample included 53 women
and 47 men. The walkthrough condition included 28 men and 22 women, whereas the
static-slide condition included 19 men and 31 women.

The conditions were similar in baseline supportive intentions, although the
walkthrough group included more men than the slide group. The mean baseline
supportive-intention index was 4.87 ($SD=0.70$) in the walkthrough condition
and 4.76 ($SD=0.53$) in the static-slide condition.

\subsection{Measures}

\subsubsection{Narrative Engagement}

Narrative engagement was the primary outcome and was measured using the
12-item Narrative Engagement Scale developed by Busselle and
Bilandzic~\cite{busselle2009engagement}. Participants rated each item from
1 (strongly disagree) to 7 (strongly agree). The scale comprises four
dimensions: narrative understanding, attentional focus, narrative presence, and
emotional engagement, with three items per dimension. The first six items were
reverse-scored, and an overall narrative-engagement score was calculated as the
mean of all 12 items. Dimension scores were examined descriptively.

\subsubsection{State Empathy}

State empathy was the secondary outcome and was assessed using Shen's 12-item
State Empathy Scale~\cite{shen2010empathy}. The scale comprises affective,
cognitive, and associative empathy. Participants rated each item from
0 (not at all) to 4 (completely) with reference to the protagonist depicted in
the presentation. An overall state-empathy score was calculated as the mean of
the 12 items, while the three dimension scores were examined descriptively.

\subsubsection{Supportive Intentions}

Four researcher-developed items assessed intentions to support
height-accessible public facilities, workplace accommodation, the challenging
of infantilising comments, and further learning about accessibility.
Participants rated the items from 1 (strongly disagree) to 7 (strongly agree)
before and after viewing the presentation for their condition.

An exploratory mean index was calculated across the four items and analysed
alongside the item-level outcomes. Because the items were researcher-developed
and represented different forms of intended support, the index was interpreted
cautiously. These items measured stated intentions rather than observed
behaviour.

\subsubsection{Manipulation Checks}

Four post-presentation items assessed perceived visual dynamism, continuity of
movement, spatial clarity, and information quantity. The first three items
examined whether participants perceived the expected presentation differences
between the continuous walkthrough and the static-slide sequence. The
information-quantity item assessed whether the two formats were perceived as
providing similar amounts of content. These items were treated as exploratory
checks on the complete presentation formats rather than as outcome measures.

\subsubsection{Open-Ended Responses}

Three open-ended questions asked participants to identify environmental and
social barriers, suggest possible changes, and comment on whether the narrative
appeared unrealistic, stereotypical, or incomplete. Ten participants provided
usable responses. These responses were treated as exploratory qualitative
material rather than as a validated measure of learning.

\subsection{Procedure}

Participants were recruited sequentially into the two conditions. The first 50
participants completed the walkthrough-video condition, after which a further
50 participants completed the static-slide condition. After consent,
participants completed demographic questions, background items, and baseline
supportive-intention ratings.

Participants in the walkthrough condition viewed the complete first-person
video, whereas participants in the slide condition advanced the full-screen
presentation at their own pace.

Sessions were conducted in a laboratory using an iPad with its built-in
speakers. The slide condition did not include narration or environmental audio.
Participants viewed the assigned stimulus without a researcher present. Before
accessing the posttest questionnaire, video participants confirmed that they had
watched the complete recording, while slide participants confirmed that they had
reached the final slide. Slide-viewing duration was not recorded, and
participants received no compensation.

Immediately afterward, all participants completed the Narrative Engagement
Scale, State Empathy Scale, posttest supportive-intention items, manipulation
checks, and were invited to answer three optional open-ended questions.

The study was approved by the Institutional Review Board of the Hong Kong Polytechnic University.

\subsection{Analysis}

Narrative engagement and state empathy were compared between conditions using
Welch's independent-samples $t$-tests. We report group means and standard
deviations, mean differences with 95\% confidence intervals, $p$-values,
and Hedges' $g$. Narrative-engagement and state-empathy subscales were examined
exploratorily.

Supportive-intention responses were paired by corresponding response order
within each condition. Within each condition, the pretest and posttest forms
were completed in a fixed one-to-one sequence, and their exported response
order was preserved. The first pretest response therefore corresponded to the
first posttest response, and so forth. No participant identifier was used;
accordingly, the supportive-intention analyses were treated as exploratory. For
each item, pre-to-post change within each condition was examined using
paired-samples $t$-tests. Differences in change between the two conditions were
tested using Welch's independent-samples $t$-tests on individual change scores.
We report pretest and posttest means, mean changes, differences in change with
95\% confidence intervals, and Hedges' $g$. The overall-index analysis was
exploratory and unadjusted. Holm correction was applied across the four
item-level between-condition comparisons.

The open-ended responses were summarised descriptively. Responses were grouped
into three broad categories: identified barriers, suggested structural or
social changes, and concerns about representation. Because only 10 usable
responses were available, the analysis was exploratory and was not intended to
establish thematic saturation. Responses were reviewed in their original
Chinese and translated into English for reporting by a bilingual member of the
research team.

\section{Results}

All 100 participants provided complete responses to the closed-ended posttest
measures, with 50 participants in each condition. Internal-consistency
estimates were $\alpha=.92$ for the walkthrough and $.98$ for the slides on
narrative engagement, and $\alpha=.84$ and $.98$, respectively, on state
empathy.

\subsection{Narrative Engagement}

Participants in the walkthrough-video condition reported greater overall
narrative engagement ($M=5.54$, $SD=0.51$) than participants in the
static-slide condition ($M=4.24$, $SD=0.73$). The mean difference was 1.30
points, 95\% CI [1.05, 1.55], Welch's $t(87.20)=10.34$, $p<.001$, with a
large standardised difference, Hedges' $g=2.05$. The result was therefore
consistent with H1.

The same pattern was observed across all four dimensions of narrative
engagement. Differences were largest for emotional engagement ($g=2.38$) and
narrative presence ($g=2.12$), followed by narrative understanding and
attentional focus (both $g=1.50$).

\subsection{State Empathy}

Overall state empathy was also higher in the walkthrough-video condition
($M=3.15$, $SD=0.33$) than in the static-slide condition ($M=2.21$,
$SD=0.60$). The mean difference was 0.94 points, 95\% CI [0.74, 1.13],
Welch's $t(76.53)=9.64$, $p<.001$, Hedges' $g=1.91$. This result was
consistent with H2.

All three empathy dimensions showed the same directional pattern. The largest
differences were found for cognitive empathy ($g=2.17$) and associative
empathy ($g=1.99$). The difference in affective empathy was smaller but
remained substantial ($g=1.14$).

\begin{table*}[!htbp]
  \caption{Main outcome comparisons}
  \label{tab:primary-results}
  \centering
  \scriptsize
  \begin{tabularx}{\textwidth}{@{}
    >{\raggedright\arraybackslash}X
    r r
    >{\raggedright\arraybackslash}p{0.20\textwidth}
    r r@{}}
    \toprule
    Outcome & Video, $M$ ($SD$) & Slides, $M$ ($SD$)
      & Difference [95\% CI] & Hedges' $g$ & $p$ \\
    \midrule
    Narrative engagement
      & 5.54 (0.51) & 4.24 (0.73)
      & 1.30 [1.05, 1.55] & 2.05 & $<.001$ \\
    \quad Narrative understanding
      & 5.37 (0.72) & 4.23 (0.78)
      & 1.13 [0.84, 1.43] & 1.50 & $<.001$ \\
    \quad Attentional focus
      & 5.35 (0.74) & 4.24 (0.73)
      & 1.11 [0.82, 1.41] & 1.50 & $<.001$ \\
    \quad Narrative presence
      & 5.55 (0.46) & 4.19 (0.77)
      & 1.36 [1.11, 1.61] & 2.12 & $<.001$ \\
    \quad Emotional engagement
      & 5.88 (0.51) & 4.29 (0.78)
      & 1.59 [1.32, 1.85] & 2.38 & $<.001$ \\
    \addlinespace
    State empathy
      & 3.15 (0.33) & 2.21 (0.60)
      & 0.94 [0.74, 1.13] & 1.91 & $<.001$ \\
    \quad Affective empathy
      & 2.82 (0.45) & 2.20 (0.61)
      & 0.61 [0.40, 0.83] & 1.14 & $<.001$ \\
    \quad Cognitive empathy
      & 3.34 (0.42) & 2.23 (0.59)
      & 1.11 [0.91, 1.32] & 2.17 & $<.001$ \\
    \quad Associative empathy
      & 3.29 (0.41) & 2.21 (0.64)
      & 1.08 [0.86, 1.29] & 1.99 & $<.001$ \\
    \bottomrule
  \end{tabularx}
  \vspace{2pt}

  \begin{minipage}{\textwidth}
    \scriptsize\emph{Note.} Each condition included 50 participants.
    Differences were calculated as walkthrough video minus static slides.
    Subscale analyses were exploratory.
  \end{minipage}
\end{table*}

\subsection{Supportive Intentions}

Baseline supportive-intention ratings were broadly similar between conditions.
As an exploratory summary across the four items, the walkthrough-video group
increased from $M=4.87$ ($SD=0.70$) at pretest to $M=6.24$ ($SD=0.55$) at
posttest. Its mean change was 1.38 points, 95\% CI [1.19, 1.56],
$t(49)=15.30$, $p<.001$. The slide group decreased from $M=4.76$
($SD=0.53$) to $M=4.55$ ($SD=0.47$), with a mean change of $-0.21$,
95\% CI [$-0.40$, $-0.01$], $t(49)=-2.12$, $p=.039$. The difference
between the two mean changes was 1.58 points, 95\% CI [1.32, 1.84],
Welch's $t(97.46)=11.96$, $p<.001$, Hedges' $g=2.37$.

Item-level analyses showed the same pattern. Ratings increased by 1.26 to 1.54
points in the walkthrough-video condition, whereas changes in the slide
condition ranged from $-0.28$ to $-0.06$. Between-condition differences in
change were significant for all four items after Holm correction. The largest
difference concerned intention to learn more about accessibility barriers.
These findings concern self-reported intentions and should not be interpreted
as observed supportive behaviour.

\begin{table}[H]
  \caption{Pretest, posttest, and change-score results for supportive intentions}
  \label{tab:supportive-intentions}
  \centering
  \scriptsize
  \begin{tabularx}{\textwidth}{@{}
    >{\raggedright\arraybackslash}X
    r r r
    r r r
    >{\raggedright\arraybackslash}p{0.15\textwidth}
    r r@{}}
    \toprule
    Item
      & \multicolumn{3}{c}{Walkthrough video}
      & \multicolumn{3}{c}{Static slides}
      & Difference in change [95\% CI]
      & $g$
      & Holm $p$ \\
    \cmidrule(lr){2-4}\cmidrule(lr){5-7}
      & Pre & Post & Change
      & Pre & Post & Change
      & & & \\
    \midrule

    Accessible public facilities
      & 4.80 (0.78) & 6.14 (0.64) & 1.34
      & 4.80 (0.57) & 4.52 (0.50) & $-0.28$
      & 1.62 [1.29, 1.95]
      & 1.94 & $<.001$ \\

    Workplace adjustments
      & 4.92 (0.83) & 6.18 (0.66) & 1.26
      & 4.72 (0.67) & 4.44 (0.58) & $-0.28$
      & 1.54 [1.22, 1.86]
      & 1.87 & $<.001$ \\

    Challenging infantilising comments
      & 4.84 (0.87) & 6.20 (0.61) & 1.36
      & 4.68 (0.71) & 4.62 (0.60) & $-0.06$
      & 1.42 [1.08, 1.76]
      & 1.66 & $<.001$ \\

    Further accessibility learning
      & 4.90 (0.93) & 6.44 (0.54) & 1.54
      & 4.82 (0.80) & 4.62 (0.49) & $-0.20$
      & 1.74 [1.39, 2.09]
      & 1.96 & $<.001$ \\

    \bottomrule
  \end{tabularx}
  \vspace{2pt}

  \begin{minipage}{\textwidth}
    \scriptsize\emph{Note.} Pretest and posttest values are $M$ ($SD$).
    Change was calculated as posttest minus pretest. Differences in change were
    calculated as walkthrough video minus static slides. Holm correction was
    applied across the four change-score comparisons.
  \end{minipage}
\end{table}

\subsection{Manipulation Checks}

Participants rated the walkthrough as more visually dynamic, continuous, and
spatially clear than the slides, with large differences across all three checks
($g=3.51$--$3.77$, all $p<.001$). They also perceived the walkthrough as
containing more information ($M=6.14$, $SD=0.86$) than the slides ($M=3.06$,
$SD=1.08$), mean difference $=3.08$, 95\% CI [2.69, 3.47], $g=3.14$,
$p<.001$. The main outcomes should therefore be interpreted as differences
between the complete presentation formats rather than as effects of movement or
continuity alone.

\subsection{Open-Ended Responses}

Ten participants provided usable open-ended responses. Of the 10 respondents,
5 had viewed the walkthrough and 5 had viewed the static slides. All identified
both physical barriers, such as inaccessible controls, storage, transport
features, or counters, and social barriers, particularly employment bias,
infantilisation, and misunderstanding. Common suggestions included lowered or
adjustable facilities, more accessible transport, and recruitment based on
ability rather than appearance.

Representational concerns were also prominent. Nine participants described the
narrative as strongly centred on difficulty or rejection. Two participants also
explicitly cautioned that a single fictionalised narrative could not represent
the experiences of all people with dwarfism. The responses therefore supported
the narrative's focus on
environmental and social barriers while also identifying a risk of
hardship-centred representation.

\subsection{Sensitivity Analyses}

As a conservative data-quality check, exact straight-lining was defined as
selecting the same raw response for every item of either 12-item scale. Seven
video-condition participants and 20 slide-condition participants met this
criterion on at least one scale.

After excluding these responses, the narrative-engagement difference remained
large: video $M=5.57$ ($SD=0.51$, $n=43$) and slides $M=4.21$
($SD=0.76$, $n=30$), mean difference $=1.36$, 95\% CI [1.04, 1.68],
Welch's $t(46.76)=8.55$, $p<.001$, $g=2.16$. The state-empathy difference
also remained large: video $M=3.17$ ($SD=0.35$) and slides $M=2.21$
($SD=0.60$), mean difference $=0.96$, 95\% CI [0.71, 1.21],
Welch's $t(42.92)=7.84$, $p<.001$, $g=2.01$.

\section{Discussion}

The walkthrough-video condition was associated with substantially higher
narrative engagement, state empathy, and supportive intentions than the
static-slide condition. However, the conditions differed in movement, temporal
continuity, pacing, participant control, narration, environmental audio, and
perceived information quantity: the video followed a fixed 10~min 52~s sequence
with narration and environmental audio, whereas participants advanced the
silent slides at their own pace. The manipulation checks further showed that
the walkthrough was perceived as more spatially clear and information-rich. The
findings should therefore be interpreted as differences between two complete
presentation formats rather than as the isolated effect of any single media
feature.

\subsection{Narrative Engagement}

For RQ1, participants who viewed the walkthrough reported greater narrative
engagement than those who viewed the slides, with a large overall difference
($g=2.05$). The pattern was consistent across narrative understanding,
attentional focus, narrative presence, and emotional engagement. The largest
differences occurred for emotional engagement and narrative presence,
suggesting that the walkthrough more strongly sustained involvement in the
protagonist's situation and the unfolding narrative.

This pattern is consistent with accounts of narrative engagement and
transportation, which emphasise attention, temporal coherence, emotional
involvement, and the construction of a continuous story world
\cite{busselle2009engagement,green2000transportation}. It is also compatible
with evidence that dynamic representations can be advantageous when movement
and temporal change are important to the material
\cite{hoeffler2007animation}. Continuous approach, reaching, object movement,
character reactions, and environmental transitions may have made the connected
barriers easier to follow than selected static frames.

Nevertheless, the difference cannot be attributed to movement alone. The video
also imposed a continuous temporal sequence, whereas slide participants
controlled how quickly they progressed through the material. The result may
therefore reflect a combination of dynamic visual information, externally
guided pacing, spatial continuity, and greater perceived information. The study
does not establish that a first-person video would always outperform a
self-paced or more information-dense slide presentation.

\subsection{State Empathy}

For RQ2, state empathy was higher after the walkthrough than after the slide
presentation ($g=1.91$). All three dimensions showed the same pattern, although
the differences were larger for cognitive and associative empathy than for
affective empathy. The walkthrough may therefore have particularly supported
understanding the protagonist's perspective and connecting the depicted events
to the protagonist's situation.

The result is consistent with earlier perspective-taking research showing
positive effects on some empathy outcomes, particularly cognitive empathy,
while also reflecting the mixed nature of the wider evidence
\cite{herrera2018empathy,vanloon2018empathy}. A meta-analysis by Ventura et al.
similarly found more consistent effects for perspective-taking than for empathy
itself~\cite{ventura2020meta}. The present findings show that the
two-dimensional walkthrough condition was
associated with higher immediate state empathy than the slide condition.

However, the video differed from the slides in temporal guidance, movement,
information quantity, and participant control. It is therefore unclear which
features contributed most strongly to the empathy difference. The findings also
do not show that the walkthrough reproduced immersive or embodied VR.
Participants did not use an HMD, control the viewpoint, interact with the
environment, or experience virtual body ownership. The results concern an
immediate response to one recorded narrative and do not establish enduring
empathy, attitude change, or accurate understanding of lived experience.

\subsection{Supportive Intentions}

For RQ3, supportive intentions increased in the walkthrough condition but
changed little overall in the slide condition. The mean increase across the four
items was 1.38 points for the walkthrough and $-0.21$ points for the slides,
producing a large between-condition difference in change ($g=2.37$). The same
direction was observed for support for accessible facilities, workplace
adjustments, challenging infantilising comments, and further accessibility
learning.

These findings suggest that the walkthrough was associated not only with
stronger engagement and empathy but also with greater immediate willingness to
support accessibility-related responses. However, the items were
researcher-developed and measured stated intentions rather than behaviour.
Changes in intentions generally translate into smaller changes in subsequent
behaviour~\cite{webb2006intentions}. The findings should therefore be treated as
exploratory evidence of an immediate motivational response rather than evidence
of behavioural change.

The small negative change in the slide condition should also be interpreted
cautiously. It does not demonstrate that static presentations reduce supportive
intentions. It may reflect sampling variation, response recalibration, or
differences in how participants navigated and paced the slide presentation.

\subsection{Format and Representational Implications}

The manipulation checks clarify the scope of the findings. Participants rated
the walkthrough as more dynamic, continuous, spatially clear, and
information-rich than the slides. In addition, the walkthrough imposed a fixed
narrative pace, whereas slide participants controlled progression through the
27 frames. The format may therefore have helped viewers connect physical
reachability, public infantilisation, and employment discrimination as parts of
an accumulating narrative, but the study cannot separate this effect from
differences in information quantity, exposure time, pacing, or participant
control.

Future comparisons should record viewing duration and distinguish between
fixed-timed slides, self-paced slides, and continuous video. More tightly
controlled studies could also vary motion, viewpoint, narration, event coverage,
and temporal pacing separately. This would help determine whether the present
effects arose from dynamic representation, guided narrative progression, richer
information, or their combination.

Practically, the results indicate that a recorded walkthrough can retain useful
features of a VR-authored narrative when HMD delivery is impractical. This may
provide a lower-resource format for classrooms, online communication, and
public engagement. The appropriate conclusion is not that screen-based video is
equivalent to VR, but that this walkthrough was associated with stronger
immediate responses than the self-paced slide presentation used in this study.

Finally, large engagement and empathy effects do not demonstrate that the
narrative authentically represents living with dwarfism or reduces
stereotyping. Disability-simulation research warns that short simulations can
increase distress or vulnerability-based perceptions and may produce limited
improvements in inclusive attitudes
\cite{narioredmond2017crip,ma2022simulation}. The fictionalised and
barrier-focused framing used here was intended to avoid claiming that viewers
temporarily ``became'' a person with dwarfism. The open-ended responses
indicated that participants recognised both environmental barriers and
discriminatory social assumptions. However, they also identified a
representational limitation: the narrative concentrated on difficulty and
rejection while providing relatively little attention to agency, achievement,
relationships, or positive everyday life. This reinforces the need for future
narratives to be developed with people of short stature and to represent a
broader range of experiences rather than barriers alone. Future development
should involve people of short stature in co-design and
evaluation~\cite{bandukda2022codesign} and should assess structural
understanding, stereotypes, and longer-term behavioural outcomes.

\section{Limitations and Future Work}

This study compared two complete presentation formats rather than isolating a
single media feature. The walkthrough differed from the slides in movement,
temporal continuity, pacing, participant control, narration, environmental
audio, and perceived information quantity; the observed differences therefore
cannot be attributed to motion or first-person continuity alone. Participants
viewed a two-dimensional recording rather than an interactive HMD experience,
and the outcomes were immediate self-reports. Because condition allocation
followed recruitment phase rather than random assignment, between-group
differences may also reflect temporal, recruitment, or other unmeasured group
differences, limiting causal inference. The convenience sample,
researcher-developed supportive-intention items, uniform responding in some
questionnaires, and the low open-ended response rate limit generalisability and
interpretation. The very high slide-condition alpha estimates may partly
reflect the greater frequency of uniform responding in that condition. The 10
qualitative responses were insufficient for thematic saturation or reliable
comparison between conditions.

Future work should compare walkthrough, slide, and HMD formats while controlling
event coverage and perceived information more closely. Studies should also use
delayed, behavioural, and qualitative measures and involve people with dwarfism
in participatory design and evaluation. Broader narratives could represent not
only barriers and discrimination but also agency, adaptation, community, and
diverse everyday experiences.

\section{Conclusion}

This study compared a 10~min 52~s first-person walkthrough video of
\emph{Big World, Little Me} with a corresponding self-paced static-slide
presentation in a non-randomised between-subjects study with 100 participants.
Participants who viewed the walkthrough reported substantially higher narrative
engagement and state empathy, while exploratory analyses also showed larger
pre-to-post increases in supportive intentions concerning accessible
facilities, workplace accommodation, challenging infantilising comments, and
further accessibility learning.

The findings suggest that recorded walkthroughs may offer a practical way to
disseminate VR-authored accessibility narratives when HMD delivery is
impractical. However, the walkthrough was also perceived as more dynamic,
continuous, spatially clear, and information-rich. The findings should
therefore be interpreted as differences between the two complete presentation
conditions, not as evidence of effects caused by motion or continuity alone.
Such media should also remain clearly framed as
fictionalised representations of selected barriers, not as equivalents to
immersive VR or to the lived experiences of people with dwarfism.

\section*{ACKNOWLEDGMENTS}

GPT-5.6 Thinking (OpenAI) was used to assist with statistical analysis,
manuscript restructuring, and language editing. The authors reviewed and
verified all AI-assisted outputs and remain responsible for the final content.

\end{document}